\documentstyle[12pt,epsf]{article}
\begin{document}
\begin{titlepage}
\rightline{\vbox{\halign{&#\hfil\cr
&UQAM-PHE-97/01\cr
&CUMQ/HEP 95\cr
&\today\cr}}}
\vspace{0.5in}
\begin{center}
{\bf SUPERSYMMETRIC QCD FLAVOUR CHANGING TOP QUARK DECAY} 
\\
\medskip
\vskip0.5in

\normalsize {{\bf G. Couture}$^{\rm a}$,
{\bf M. Frank}$^{\rm b}$, 
and {\bf H. K\"{o}nig}$^{\rm a,b}$}
\smallskip
\medskip

{ \sl $^a$
D\'{e}partement de Physique, Universit\'{e} du Qu\'{e}bec \`{a} Montr\'{e}al\\ 
C.P. 8888, Succ. Centre Ville, Montr\'eal, Qu\'{e}bec, Canada, H3C 3P8\\
$^b$Department of Physics, Concordia University, 
1455 De Maisonneuve Blvd. W.\\ Montr\'eal, Quebec, Canada, H3G 1M8 
}
\smallskip
\end{center}
\vskip1.0in

\noindent{\large\bf Abstract}
\smallskip

We present a detailed and complete calculation of the gluino
and scalar quarks contribution to the flavour-changing
top quark decay into a charm quark and a photon, gluon,
or a $Z^0$\ boson within the minimal supersymmetric standard model 
including flavour changing gluino-quarks-scalar quarks couplings in the 
right-handed sector. We compare the results with the ones
presented in an earlier paper where we considered flavour
changing couplings only in the left-handed sector.
We show that these new couplings have important consequences
leading to a large enhancement when the mixing of the scalar
partners of the left- and right-handed top quark is included.
Furthermore CP violation in the flavour changing top
quark decay will occur when a SUSY phase is taken into account.

\end{titlepage}

\baselineskip=20pt

\newpage
\pagenumbering{arabic}

\section{\bf Introduction}
\label{intro}

Flavour changing top quark decay modes are  a promising test ground
for models beyond the standard model (SM). While in the SM
the branching ratios of the decays $t\rightarrow c\gamma,\ cg$\
and $cZ$\ are far away from experimental reach 
~\cite{dc}-\cite{ehs}, the authors of ~\cite{ehs}-\cite{ggk}
showed that they are enhanced by several
(3-4) orders of magnitude in Two-Higgs-doublet models (THDM's).

Nowadays, CDF ~\cite{cdf1}-\cite{cdf2} and $D\emptyset$ ~\cite{d0}
have begun to explore flavour changing top quark decays and 
interesting bounds have been reported ~\cite{cdf2}. A 
systematic examination of anomalous top quark quark interactions
is actively pursued ~\cite{hkyz1}-\cite{hwy}.

Within supersymmetry, the decays $t\rightarrow cV$\ were first
considered in ~\cite{loy} and the authors obtained the same
enhancement as in the THDM's. However as we have pointed out
in a recent paper ~\cite{chk}, in their calculation 
of the QCD corrections they had
an inconsistency basically due to the lack of gauge invariance
arising from the omission of the gluino-gluino-gluon coupling.
They also did not include the non-negligible mixing of the scalar
partners of the left and right handed quarks.
In a very recent paper ~\cite{lnr} the calculations were
redone for the weak sector with charginos and neutralinos
within the relevant loops including the mixing of the 
scalar quarks, where it was shown that supersymmetric
contributions to $t\rightarrow cV$\ can be up to 5 orders
of magnitude larger than their SM counterparts.
 
In our previous paper ~\cite{chk} 
the results of the gluino and scalar quarks contribution to the
flavour-changing top quark decay into a charm quark and a photon,
a gluon or a $Z^0$\ boson within the minimal supersymmetric 
standard model (MSSM) were presented. 
We included the mixing of the scalar
partners of the left- and right-handed top quark and showed 
that it has several effects, the most important of which is to
greatly enhance the cZ decay mode for large values of the soft
SUSY-breaking scalar mass $m_S$\ and to give rise to a GIM-like
suppression in the $c\gamma$\ mode for certain combinations
of parameters.

However the analysis of ~\cite{chk} considered 
flavour-changing strong interactions
between the gluino, the quarks and their scalar partners only in
the left-handed sector and kept the right-handed sector flavour-diagonal.
This is a common assumption within the MSSM (see ~\cite{chk}
and references therein) and might not be necessarily the case in any kind
of extension of the MSSM, or more general assumptions within the MSSM.

The goal of this paper is to recalculate the flavour-changing top 
quark decays 
including flavour-changing couplings within the right-handed sector.
We will assume maximal flavour-changing in both sectors and analyse  
how the previous results will be changed. Furthermore we show that
flavour-changing couplings in the left- and right-handed sectors 
lead to a CP violating term proportional to the gluino mass for the
top quark decay modes under consideration, which will be investigated
in a further paper ~\cite{cfk}.

The Feynman diagrams and the couplings leading to the decay
modes $t\rightarrow c\gamma,\ cZ\ {\rm and}\ cg$\ as well
as the mass matrix of the scalar top quark are given in
~\cite{chk}. The only difference will be the flavour-changing
gluino-scalar quarks-quarks coupling in Eq.(6) of ~\cite{chk}, which
will be taken in the present paper in the most general way:

\begin{equation}
\label{fceq}
{\cal L_{FC}}=-\sqrt{2}g_sT^a\overline{\tilde g_a}
\biggl\lbrack K^{\tilde g}_L(c_\Theta\tilde
q_1-s_\Theta\tilde q_2)P_L-K^{\tilde g}_R(s_\Theta\tilde q_1+
c_\Theta\tilde q_2)P_R\biggr\rbrack q+h.c.
\end{equation}

Here $K^{\tilde g}_{L,R}$\
 is the supersymmetric version of the Kobayashi--Maskawa
matrix and $\Theta$\ is the mixing angle of the scalar
partners of the left- and right-handed quarks. 
$\tilde q_1$\ and $\tilde q_2$\ are the
mass eigenstates which are related to the current eigenstates
$\tilde q_L$\  and $\tilde q_R$\ by

\begin{equation}
\label{mascur}
\tilde q_1=\cos\Theta\tilde q_L+\sin\Theta q_R,\quad
\tilde q_2=-\sin\Theta\tilde q_L+\cos\Theta q_R
\end{equation} 

\section{\bf SUSY QCD flavour changing top quark decay}
\label{sqftd}

After summation over all diagrams, we obtain the following
effective $tcV$ vertex, neglecting the charm quark mass:

\begin{eqnarray}
\label{geneq}
M^\alpha_{\mu V}&=&-i{{\alpha_s}\over{2\pi}}
\overline u_{p_2}\biggl\lbrack
\gamma_\mu \Bigl(P_L V_{VL}^\alpha+P_R V_{VR}^\alpha\Bigr)
+{{P_\mu}\over{m_{\rm top}}}\Bigl(P_L T_{VL}^\alpha
+P_RT_{VR}^\alpha\Bigr)\biggl\rbrack u_{p_1}\\
V_{\gamma L}^\alpha&=&ee_qC_2(F)\biggl\lbrace 
K^{\ast\tilde g}_{\alpha 2L}K^{\tilde g}_{\alpha 3L}
\Bigl\lbrack
c_{\Theta_\alpha}^2 (C^{11\alpha}_\epsilon-C^{1\alpha}_{SE})
+s_{\Theta_\alpha}^2(C^{22\alpha}_\epsilon-C^{2\alpha}_{SE})
\Bigr\rbrack\nonumber\\
& &+K^{\ast\tilde g}_{\alpha 2L}K^{\tilde g}_{\alpha 3R}
c_{\Theta_\alpha}s_{\Theta_\alpha}{{m_{\tilde g}}\over{m_{\rm top}}}
\Bigl\lbrack C^{1\alpha}_{SEG}-C^{1\alpha}_{SEG}\vert_{m_{\rm top}^2=0}
-C^{2\alpha}_{SEG}+C^{2\alpha}_{SEG}\vert_{m_{\rm top}^2=0}\Bigr\rbrack
\biggr\rbrace
\nonumber\\
V_{\gamma R}^\alpha&=&ee_qC_2(F)\biggl\lbrace 
K^{\ast\tilde g}_{\alpha 2R}K^{\tilde g}_{\alpha 3R}
\Bigl\lbrack
s_{\Theta_\alpha}^2 (C^{11\alpha}_\epsilon-C^{1\alpha}_{SE})
+c_{\Theta_\alpha}^2(C^{22\alpha}_\epsilon-C^{2\alpha}_{SE})
\Bigr\rbrack\nonumber\\
& &+K^{\ast\tilde g}_{\alpha 2R}K^{\tilde g}_{\alpha 3L}
c_{\Theta_\alpha}s_{\Theta_\alpha}{{m_{\tilde g}}\over{m_{\rm top}}}
\Bigl\lbrack C^{1\alpha}_{SEG}-C^{1\alpha}_{SEG}\vert_{m_{\rm top}^2=0}
-C^{2\alpha}_{SEG}+C^{2\alpha}_{SEG}\vert_{m_{\rm top}^2=0}\Bigr\rbrack
\biggr\rbrace
\nonumber\\
T_{\gamma L}^\alpha&=&ee_qC_2(F)\biggl\lbrace
K^{\ast\tilde g}_{\alpha 2R}K^{\tilde g}_{\alpha 3R}
\Bigl\lbrack s_{\Theta_\alpha}^2
C^{11\alpha}_{\rm top}+c_{\Theta_\alpha}^2 C^{22\alpha}_{\rm top}
\Bigr\rbrack
-K^{\ast\tilde g}_{\alpha 2R}K^{\tilde g}_{\alpha 3L}
c_{\Theta_\alpha}s_{\Theta_\alpha}\Bigl\lbrack
C^{11\alpha}_{\tilde g\rm top}-C^{22\alpha}_{\tilde g\rm top}
\Bigr\rbrack\biggr\rbrace
\nonumber\\
T_{\gamma R}^\alpha&=&ee_qC_2(F)\biggl\lbrace
K^{\ast\tilde g}_{\alpha 2L}K^{\tilde g}_{\alpha 3L}
\Bigl\lbrack c_{\Theta_\alpha}^2
C^{11\alpha}_{\rm top}+s_{\Theta_\alpha}^2 C^{22\alpha}_{\rm top}
\Bigr\rbrack
-K^{\ast\tilde g}_{\alpha 2L}K^{\tilde g}_{\alpha 3R}
c_{\Theta_\alpha}s_{\Theta_\alpha}\Bigl\lbrack
C^{11\alpha}_{\tilde g\rm top}-C^{22\alpha}_{\tilde g\rm top}
\Bigr\rbrack\biggr\rbrace
\nonumber\\
V_{gL}^\alpha&=&g_sT^a\biggl\lbrace
K^{\ast\tilde g}_{\alpha 2L}K^{\tilde g}_{\alpha 3L}
\Bigl\lbrace\lbrack -{1\over 2}C_2(G)+C_2(F)
\rbrack\lbrack c_{\Theta_\alpha}^2 C^{11\alpha}_\epsilon+
s_{\Theta_\alpha}^2 C^{22\alpha}_\epsilon\rbrack-C_2(F)\lbrack
c_{\Theta_\alpha}^2 C^{1\alpha}_{SE}+s_{\Theta_\alpha}^2
C^{2\alpha}_{SE}\rbrack\nonumber\\
& &+{1\over 2}C_2(G)\Bigl\lbrack c_{\Theta_\alpha}^2\lbrack
C^{\tilde g 1\alpha}_\epsilon+C^{1\alpha}_{\tilde g}
+C_{q^2}^{1\alpha}+C_t^{1\alpha}\rbrack
+s_{\Theta_\alpha}^2\lbrack
C^{\tilde g 2\alpha}_\epsilon+C^{2\alpha}_{\tilde g}
+C_{q^2}^{2\alpha}+C_t^{2\alpha}
\rbrack\Bigr\rbrack\Bigr\rbrace\nonumber\\
& &+K^{\ast\tilde g}_{\alpha 2L}K^{\tilde g}_{\alpha 3R}
c_{\Theta_\alpha}s_{\Theta_\alpha}\Bigl\lbrack
C_2(F){{m_{\tilde g}}\over{m_{\rm top}}}
\Bigl\lbrack C^{1\alpha}_{SEG}-C^{1\alpha}_{SEG}\vert_{m_{\rm top}^2=0}
-C^{2\alpha}_{SEG}+C^{2\alpha}_{SEG}\vert_{m_{\rm top}^2=0}\Bigr\rbrack
\nonumber\\
& &-{1\over 2}C_2(G){{m_{\rm top}}\over{m_{\tilde g}}}
\lbrack C^{1\alpha}_{\tilde g}
-C^{2\alpha}_{\tilde g}\rbrack\Bigr\rbrack\biggr\rbrace
\nonumber\\
V_{gR}^\alpha&=&g_sT^a\biggl\lbrace
K^{\ast\tilde g}_{\alpha 2R}K^{\tilde g}_{\alpha 3R}
\Bigl\lbrace\lbrack -{1\over 2}C_2(G)+C_2(F)
\rbrack\lbrack s_{\Theta_\alpha}^2 C^{11\alpha}_\epsilon+
c_{\Theta_\alpha}^2 C^{22\alpha}_\epsilon\rbrack-C_2(F)\lbrack
s_{\Theta_\alpha}^2 C^{1\alpha}_{SE}+c_{\Theta_\alpha}^2
C^{2\alpha}_{SE}\rbrack\nonumber\\
& &+{1\over 2}C_2(G)\Bigl\lbrack s_{\Theta_\alpha}^2\lbrack
C^{\tilde g 1\alpha}_\epsilon+C^{1\alpha}_{\tilde g}
+C_{q^2}^{1\alpha}+C_t^{1\alpha}\rbrack
+c_{\Theta_\alpha}^2\lbrack
C^{\tilde g 2\alpha}_\epsilon+C^{2\alpha}_{\tilde g}
+C_{q^2}^{2\alpha}+C_t^{2\alpha}
\rbrack\Bigr\rbrack\Bigr\rbrace\nonumber\\
& &+K^{\ast\tilde g}_{\alpha 2R}K^{\tilde g}_{\alpha 3L}
c_{\Theta_\alpha}s_{\Theta_\alpha}\Bigl\lbrack
C_2(F){{m_{\tilde g}}\over{m_{\rm top}}}
\Bigl\lbrack C^{1\alpha}_{SEG}-C^{1\alpha}_{SEG}\vert_{m_{\rm top}^2=0}
-C^{2\alpha}_{SEG}+C^{2\alpha}_{SEG}\vert_{m_{\rm top}^2=0}\Bigr\rbrack
\nonumber\\
& &-{1\over 2}C_2(G){{m_{\rm top}}\over{m_{\tilde g}}}
\lbrack C^{1\alpha}_{\tilde g}
-C^{2\alpha}_{\tilde g}\rbrack\Bigr\rbrack\biggr\rbrace
\nonumber\\
T_{gL}^\alpha&=&g_sT^a\biggl\lbrace
K^{\ast\tilde g}_{\alpha 2R}K^{\tilde g}_{\alpha 3R}
\Bigl\lbrace\lbrack -{1\over 2}C_2(G)+C_2(F)
\rbrack\lbrack s_{\Theta_\alpha}^2 C^{11\alpha}_{\rm top}+
c_{\Theta_\alpha}^2 C^{22\alpha}_{\rm top}\rbrack
\nonumber\\
& &-{1\over 2}C_2(G)\Bigl\lbrack s_{\Theta_\alpha}^2
C^{1\alpha}_t
+c_{\Theta_\alpha}^2
C^{2\alpha}_t
\Bigr\rbrack\Bigr\rbrace\nonumber\\
& &-K^{\ast\tilde g}_{\alpha 2R}K^{\tilde g}_{\alpha 3L}
c_{\Theta_\alpha}s_{\Theta_\alpha}\Bigl\lbrace
\lbrack -{1\over 2}C_2(G)+C_2(F)\rbrack\lbrack
 C^{11\alpha}_{\tilde g {\rm top}}- C^{22\alpha}_{\tilde g{\rm top}}
\rbrack\nonumber\\
& &+{1\over 2}C_2(G)\Bigl\lbrack C^{1\alpha}_{\tilde g t}
-C^{2\alpha}_{\tilde g t}-{{m_{\rm top}}\over{m_{\tilde g}}}
\Bigl(C^{1\alpha}_{\tilde g}
-C^{2\alpha}_{\tilde g}\Bigr)\Bigr\rbrack\Bigr\rbrace\biggr\rbrace
\nonumber\\
T_{gR}^\alpha&=&g_sT^a\biggl\lbrace
K^{\ast\tilde g}_{\alpha 2L}K^{\tilde g}_{\alpha 3L}
\Bigl\lbrace\lbrack -{1\over 2}C_2(G)+C_2(F)
\rbrack\lbrack c_{\Theta_\alpha}^2 C^{11\alpha}_{\rm top}+
s_{\Theta_\alpha}^2 C^{22\alpha}_{\rm top}\rbrack
\nonumber\\
& &-{1\over 2}C_2(G)\Bigl\lbrack c_{\Theta_\alpha}^2
C^{1\alpha}_t
+s_{\Theta_\alpha}^2
C^{2\alpha}_t
\Bigr\rbrack\Bigr\rbrace\nonumber\\
& &-K^{\ast\tilde g}_{\alpha 2L}K^{\tilde g}_{\alpha 3R}
c_{\Theta_\alpha}s_{\Theta_\alpha}\Bigl\lbrace
\lbrack -{1\over 2}C_2(G)+C_2(F)\rbrack\lbrack
 C^{11\alpha}_{\tilde g {\rm top}}- C^{22\alpha}_{\tilde g{\rm top}}
\rbrack\nonumber\\
& &+{1\over 2}C_2(G)\Bigl\lbrack C^{1\alpha}_{\tilde g t}
-C^{2\alpha}_{\tilde g t}-{{m_{\rm top}}\over{m_{\tilde g}}}
\Bigl(C^{1\alpha}_{\tilde g}
-C^{2\alpha}_{\tilde g}\Bigr)\Bigr\rbrack\Bigr\rbrace\biggr\rbrace
\nonumber\\
V_{ZL}^\alpha&=&{{e}\over{s_Wc_W}}C_2(F)\biggl\lbrace
K^{\ast\tilde g}_{\alpha 2L}K^{\tilde g}_{\alpha 3L}
\Bigl\lbrace(T_{3L}c^2_{\Theta_\alpha}-e_qs_W^2)
c_{\Theta_\alpha}^2 C^{11\alpha}_\epsilon
+(T_{3L}s^2_{\Theta_\alpha}-e_qs_W^2)
s_{\Theta_\alpha}^2 C^{22\alpha}_\epsilon\nonumber\\
& &+T_{3L}c^2_{\Theta_\alpha} s^2_{\Theta_\alpha}
(C^{12\alpha}_\epsilon+C^{21\alpha}_
\epsilon)-(T_{3L}-e_qs_W^2)\lbrack c^2_{\Theta_\alpha}
C^{1\alpha}_{SE}+s^2_{\Theta_\alpha} C^{2\alpha}_{SE}
\rbrack\Bigr\rbrace\nonumber\\
& &+K^{\ast\tilde g}_{\alpha 2L}K^{\tilde g}_{\alpha 3R}
c_{\Theta_\alpha}s_{\Theta_\alpha}(T_{3L}-e_qs_W^2)
{{m_{\tilde g}}\over{m_{\rm top}}}
\Bigl\lbrack C^{1\alpha}_{SEG}-C^{1\alpha}_{SEG}\vert_{m_{\rm top}^2=0}
-C^{2\alpha}_{SEG}+C^{2\alpha}_{SEG}\vert_{m_{\rm top}^2=0}\Bigr\rbrack
\biggr\rbrace
\nonumber\\
V_{ZR}^\alpha&=&{{e}\over{s_Wc_W}}C_2(F)\biggl\lbrace
K^{\ast\tilde g}_{\alpha 2R}K^{\tilde g}_{\alpha 3R}
\Bigl\lbrace(T_{3L}c^2_{\Theta_\alpha}-e_qs_W^2)
s_{\Theta_\alpha}^2 C^{11\alpha}_\epsilon
+(T_{3L}s^2_{\Theta_\alpha}-e_qs_W^2)
c_{\Theta_\alpha}^2 C^{22\alpha}_\epsilon\nonumber\\
& &-T_{3L}c^2_{\Theta_\alpha} s^2_{\Theta_\alpha}
(C^{12\alpha}_\epsilon+C^{21\alpha}_
\epsilon)+e_qs_W^2\lbrack s^2_{\Theta_\alpha}
C^{1\alpha}_{SE}+c^2_{\Theta_\alpha} C^{2\alpha}_{SE}
\rbrack\Bigr\rbrace\nonumber\\
& &-K^{\ast\tilde g}_{\alpha 2L}K^{\tilde g}_{\alpha 3R}
c_{\Theta_\alpha}s_{\Theta_\alpha}e_qs_W^2
{{m_{\tilde g}}\over{m_{\rm top}}}
\Bigl\lbrack C^{1\alpha}_{SEG}-C^{1\alpha}_{SEG}\vert_{m_{\rm top}^2=0}
-C^{2\alpha}_{SEG}+C^{2\alpha}_{SEG}\vert_{m_{\rm top}^2=0}\Bigr\rbrack
\biggr\rbrace
\nonumber\\
T_{ZL}^\alpha&=&{{e}\over{s_Wc_W}}C_2(F)\biggl\lbrace
K^{\ast\tilde g}_{\alpha 2R}K^{\tilde g}_{\alpha 3R}
\Bigl\lbrace
(T_{3L}c^2_{\Theta_\alpha}-e_qs_W^2)
s_{\Theta_\alpha}^2 C^{11\alpha}_{\rm top}
+(T_{3L}s^2_{\Theta_\alpha}-e_qs_W^2)
c_{\Theta_\alpha}^2 C^{22\alpha}_{\rm top}\nonumber\\
& &-T_{3L}c^2_{\Theta_\alpha} s^2_{\Theta_\alpha}
(C^{12\alpha}_{\rm top}+C^{21\alpha}_{\rm top})\Bigr\rbrace
\nonumber\\
& &-K^{\ast\tilde g}_{\alpha 2R}K^{\tilde g}_{\alpha 3L}
c_{\Theta_\alpha}s_{\Theta_\alpha}\Bigl\lbrack
(T_{3L}c^2_{\Theta_\alpha}-e_qs_W^2) C^{11\alpha}_{\tilde g{\rm top}}
-(T_{3L}s^2_{\Theta_\alpha}-e_qs_W^2)C^{22\alpha}_{\tilde g{\rm top}}
\nonumber\\
& &-T_{3L}\lbrack c^2_{\Theta_\alpha} C^{12\alpha}_{\tilde g{\rm top}}
- s^2_{\Theta_\alpha}C^{21\alpha}_{\tilde g{\rm top}}\rbrack
\Bigr\rbrack\biggr\rbrace\nonumber\\
T_{ZR}^\alpha&=&{{e}\over{s_Wc_W}}C_2(F)\biggl\lbrace
K^{\ast\tilde g}_{\alpha 2L}K^{\tilde g}_{\alpha 3L}
\Bigl\lbrace
(T_{3L}c^2_{\Theta_\alpha}-e_qs_W^2)
c_{\Theta_\alpha}^2 C^{11\alpha}_{\rm top}
+(T_{3L}s^2_{\Theta_\alpha}-e_qs_W^2)
s_{\Theta_\alpha}^2 C^{22\alpha}_{\rm top}\nonumber\\
& &+T_{3L}c^2_{\Theta_\alpha} s^2_{\Theta_\alpha}
(C^{12\alpha}_{\rm top}+C^{21\alpha}_{\rm top})\Bigr\rbrace
\nonumber\\
& &-K^{\ast\tilde g}_{\alpha 2L}K^{\tilde g}_{\alpha 3R}
c_{\Theta_\alpha}s_{\Theta_\alpha}\Bigl\lbrack
(T_{3L}c^2_{\Theta_\alpha}-e_qs_W^2) C^{11\alpha}_{\tilde g{\rm top}}
-(T_{3L}s^2_{\Theta_\alpha}-e_qs_W^2)C^{22\alpha}_{\tilde g{\rm top}}
\nonumber\\
& &+T_{3L}\lbrack s^2_{\Theta_\alpha} C^{12\alpha}_{\tilde g{\rm top}}
- c^2_{\Theta_\alpha}C^{21\alpha}_{\tilde g{\rm top}}\rbrack
\Bigr\rbrack\biggr\rbrace\nonumber\\
C^{kl\alpha}_\epsilon
&=&\int\limits_0^1d\alpha_1
\int\limits_0^{1-\alpha_1}d\alpha_2
\lbrack {1\over\epsilon}-\gamma+\ln(4\pi\mu^2)-\ln(f_{kl}^
\alpha)\rbrack\nonumber\\
 C^{kl\alpha}_{\rm top}&=&
\int\limits_0^1d\alpha_1
\int\limits_0^{1-\alpha_1}d\alpha_2
{{m^2_{\rm top}\alpha_1(1-\alpha_1-\alpha_2)}\over{ f_{kl}^ 
\alpha}}\nonumber\\
 C^{kl\alpha}_{\tilde g{\rm top}}&=&
\int\limits_0^1d\alpha_1
\int\limits_0^{1-\alpha_1}d\alpha_2
{{m_{\tilde g}m_{\rm top}(1-\alpha_1-\alpha_2)}\over{ f_{kl}^ 
\alpha}}\nonumber\\
C^{k\alpha}_{SE}&=&\int\limits_0^1 d\alpha_1\alpha_1
\lbrack {1\over\epsilon}-\gamma+\ln(4\pi\mu^2)-\ln(g_k^\alpha)
\rbrack\nonumber\\
C^{k\alpha}_{SEG}&=&{1\over\alpha_1}C^{k\alpha}_{SE}\nonumber\\
C^{\tilde g k\alpha}_\epsilon&=&\int\limits_0^1d\alpha_1
\int\limits_0^{1-\alpha_1}d\alpha_2
\lbrack {1\over\epsilon}-\gamma-1+\ln(4\pi\mu^2)-\ln(h_k^
\alpha)\rbrack\nonumber\\
C^{k\alpha}_{\tilde g}&=&\int\limits_0^1d\alpha_1
\int\limits_0^{1-\alpha_1}d\alpha_2 {{m_{\tilde g}^2}
\over{h^\alpha_k}}\nonumber\\
C_{q^2}^{k\alpha}&=&\int\limits_0^1d\alpha_1
\int\limits_0^{1-\alpha_1}d\alpha_2{{q^2\alpha_1\alpha_2}\over
{h^\alpha_k}}\nonumber\\
C^{k\alpha}_t&=&\int\limits_0^1d\alpha_1
\int\limits_0^{1-\alpha_1}d\alpha_2
{{m^2_{\rm top}\alpha_1(1-\alpha_1-\alpha_2)}\over{h_k^\alpha}}  
\nonumber\\
C^{k\alpha}_{\tilde g t}&=&\int\limits_0^1d\alpha_1
\int\limits_0^{1-\alpha_1}d\alpha_2
{{m_{\tilde g}m_{\rm top}(1-\alpha_1-\alpha_2)}\over{h_k^\alpha}}  
\nonumber\\
f_{kl}^{\alpha}&=&m^2_{\tilde g}-(m^2_{\tilde g}-
m_{\tilde q^\alpha_k}^2)\alpha_1-(m^2_{\tilde g}-
m_{\tilde q^\alpha_l}^2)\alpha_2-
m^2_{\rm top}\alpha_1(1-\alpha_1-\alpha_2)
- q^2\alpha_1\alpha_2\nonumber\\
g^\alpha_k&=&m^2_{\tilde g}-(m^2_{\tilde g}-
m_{\tilde q^\alpha_k}^2)\alpha_1-
m^2_{\rm top}\alpha_1(1-\alpha_1)\nonumber\\
h^\alpha_k&=&m_{\tilde q^\alpha_k}^2-(m_{\tilde q^\alpha_k}^2-
m^2_{\tilde g})(\alpha_1+\alpha_2)-
m^2_{\rm top}\alpha_1(1-\alpha_1-\alpha_2)
- q^2\alpha_1\alpha_2\nonumber
\end{eqnarray}

where $\epsilon = 2-d/2$,\ $C_2(F)=4/3$\ and $C_2(G)=3$\ for SU(3). 
If $\alpha\not={\rm top}$\ we have $c_{\Theta_\alpha}=1$. 
Using the spin-1 condition ($q_\mu=(p_1-p_2)_\mu=0$)\ we
can write $P_\mu=(p_1+p_2)_\mu=2p_{1\mu}$.
$K^{\tilde g}_{\alpha iL,R}$\ is the SUSY--Kobayashi--Maskawa matrix;
which, as explained in ~\cite{chk}, will be parameterized by a small 
number $\varepsilon$\  
(not to be confused with the $\epsilon$\ above) to be taken as 
$\varepsilon^2=1/4$\ ~\cite{loy,dun}.
It is straightforward at this point to verify that all divergent terms cancel
exactly in a non trivial way, without making use of the GIM mechanism.
The results of ~\cite{chk} are reproduced with 
$K^{\tilde g}_{\alpha 2,3R}=0$,
that is $V_{VR}=0=T_{VL}$. Note that with $K^{\tilde g}_{\alpha 2,3R}
\not =0$\ we obtain terms proportional to the gluino mass, which 
might become dominant for large gluino masses.

A further crucial test is also provided by the nature of the current.
Using the following identity:

\begin{equation}
\label{ident}
\overline u_{p_2}{{P^\mu}\over{m_{\rm top}}}P_{L,R} u_{p_1}
\equiv \overline u_{p_2}\lbrack\gamma_\mu P_{R,L}+
i\sigma_{\mu\nu}{{q^\mu}\over{m_{\rm top}}}P_{L,R}\rbrack u_{p_1}
\end{equation}

and after Feynman integration with:

\begin{eqnarray}
\label{feynid}
\biggl\lbrack
C^{ii\alpha}_\epsilon+C^{ii\alpha}_{\rm top}-C^{i\alpha}_
{SE}\biggr\rbrack_{q^2=0}\equiv & &0\nonumber\\
\biggl\lbrack C^{ii\alpha}_\epsilon+C^{ii\alpha}_{\rm top}
-C^{\tilde g i\alpha}_\epsilon-C^{i\alpha}_{\tilde g}
\biggr\rbrack_{q^2=0}\equiv & &0\nonumber\\
\biggl\lbrack
{{m_{\tilde g}}\over{m_{\rm top}}}\Bigl\lbrace C_{SEG}^{i\alpha}-
C_{SEG}^{i\alpha}\vert_{m^2_{\rm top}=0}\Bigr\rbrace-C^{ii\alpha}_{\tilde g
{\rm top}}\biggr\rbrack_{q^2=0}\equiv & &0\nonumber\\
\biggl\lbrack
C^{ii\alpha}_{\tilde g{\rm top}}-C^{ii\alpha}_{\tilde g t}
\biggr\rbrack_{q^2=0}\equiv & &0
\end{eqnarray}

we can show that the quantity in front of the $\gamma^\mu$ term 
vanishes in the 
limit $q^2\to 0$, as required by gauge invariance, that is 
$V_{VL,R}=-T_{VR,L}$\ for $V=\gamma, g$. For $V=Z$, that
is $q^2=m_Z^2$, the relations above do not hold anymore.
We do the first Feynman integration by hand and the
second one numerically\footnote{We think that in the computer age 
it is not necessary to present the results in the
form of the Passarino-Veltman functions, which would
make the results only more difficult to read, but refer the
interested reader to ~\cite{ddk}, where similiar calculations
have been done. See also ~\cite{lnr,ita}.}.

In a recent paper ~\cite{kon} one of us (H.K.) considered the
gluino and neutralino contributions to the direct CP violating
parameter $\epsilon'$. The Feynman diagrams and calculations
were similiar. It is straightforward to show that  eq.(\ref{geneq})
reproduce the eq.(A.9) in ~\cite{kon} by replacing $m_{\rm top}$\ with
$m_s$\ and putting the down quark there to zero.

We assumed that both couplings
of the gluino to the left- and right-handed quarks and their
superpartners are flavour non diagonal and to be of the same
order, that is we take $K^{\tilde g}_{abR}=e^{-i\Phi_S} K_{ab}$\
and $K^{\tilde g}_{abL}=e^{+i\Phi_S} K_{ab}$, where $\Phi_S$\
is a supersymmetric CP violation phase ~\cite{kon}.

In eq.(\ref{geneq}) this phase only comes in when $K_L^{\tilde g}$\
is multiplied by $K_R^{\tilde g}$\ and, as can be seen, these terms are
proportional to the gluino mass. However this SUSY CP violating
phase is strongly bounded by the electric dipole moment of the
neutron (EDMN) to be of the order of $10^{-2}-10^{-3}$, if not the
SUSY masses are heavier than several TEV's (see references given
in ~\cite{kon}).  
We are not interested here in the consequences of
this phase leading to CP violating flavour changing top
quark decay, which will be presented elsewhere ~\cite{cfk}.
In the following we put $\Phi_S=0$.

When summing over all scalar quarks within the loops,
the scalar up quark contributions cancels because of the unitarity
of $K_{ab}$, and with $K_{23}=-K_{32}$\ the mass
splitting of the scalar top quark and the scalar charm
quark comes into account. This was taken
to be $m_{\tilde c}=0.9~m_{\tilde t}$\ in ~\cite{loy}, and therefore
too small for a top quark mass of 174 GeV.
If all scalar quark masses would be the samei, the
decay rate of $t\rightarrow cV$\ would be identical
to 0. As a final result we obtain:

\begin{eqnarray}
\label{decrate}
\Gamma_S(t\rightarrow c V)&=&
{{\alpha_s^2}
\over{128\pi^3}}m_{\rm top}
\biggl( 1-{{m_{V}^2}\over{m^2_{\rm top}}} \biggr)^2\varepsilon^2
\Bigl\lbrack (V_{VL}^2+V_{VR}^2)
\Bigl( 2+{{m^2_{\rm top}}\over{m^2_{V}}} \Bigr)\\
& &
-2(V_{VL}T_{VR}+V_{VR}T_{VL})\Bigl( 1-{{m^2_{\rm top}}\over{m^2_{V}}} \Bigr)
-(T_{VL}^2+T_{VR}^2)\Bigl( 2-{{m^2_{V}}\over{m^2_{\rm top}}}-
{{m^2_{\rm top}}\over{m^2_{V}}} \Bigr) \Bigr\rbrack
\nonumber
\end{eqnarray}

where  $\displaystyle{V_{VL,R}=V_{VL,R}^{\tilde t}-
V_{VL,R}^{\tilde c}}$\
and $\displaystyle{T_{VL,R}=T_{VL,R}^{\tilde t}-
T_{VL,R}^{\tilde c}}$.
As explained above for $V=\gamma,g$ we have 
$V_{VL,R}=-T_{VR,L}$\ and
all terms containing $m_V^2$\ are absent.

We define ~\cite{ehs}:
$\displaystyle{B(t\rightarrow cV)=\Gamma_S(t\rightarrow cV)/
\Gamma_W(t\rightarrow bW^+)}$\ where 

\begin{equation}
\label{decwt}
\Gamma_W(t\rightarrow bW^+)={{\alpha}\over{16\sin^2\Theta_W}}
m_{\rm top}\Bigl(1-{{m^2_{W^+}}\over{m^2_{\rm top}}}\Bigr)^2\Bigl(2+
{{m^2_{\rm top}}\over{m^2_{W^+}}}\Bigr)
\end{equation}

Our input parameters are 
$m_{\rm top}=174$\ GeV and the strong coupling
constant $\alpha_s=1.4675/\ln({{m^2_{\rm top}}\over
{\Lambda^2_{\rm QCD}}})=0.107$\ with $\Lambda_{\rm QCD}=
0.18$\ GeV ~\cite{ehs}.

\section{Discussions}
\label{discu}

To compare the new results with flavour changing couplings in the
right- and left-handed sector with the ones already presented in ~\cite{chk},
where flavour changing couplings only in the left-handed sector
was considered, we present the same plots as in ~\cite{chk}.
The general discussion remains the same and we will only present 
the changes when flavour changing in the right-handed
sector is inlcuded.

In Fig. 1 we present the branching ratio
$B(t\rightarrow c Z)$ 
\ as a function of the scalar
mass $m_S$\ for a gluino mass of 100 GeV. 
We see that without mixing, the branching 
ratio decreases rapidly with increasing scalar mass
and is hardly changed when flavour changing in
the right-handed sector is included.
However the mixing has a drastic effect. It enhances the branching ratio
 by up to 4 orders of
magnitude for large $m_s$\ and is enhanced by another factor
of 5 when flavour changing occurs in both sectors. 

\begin{figure}[hbtp]
\begin{center}
\mbox{\epsfxsize=144mm\epsffile{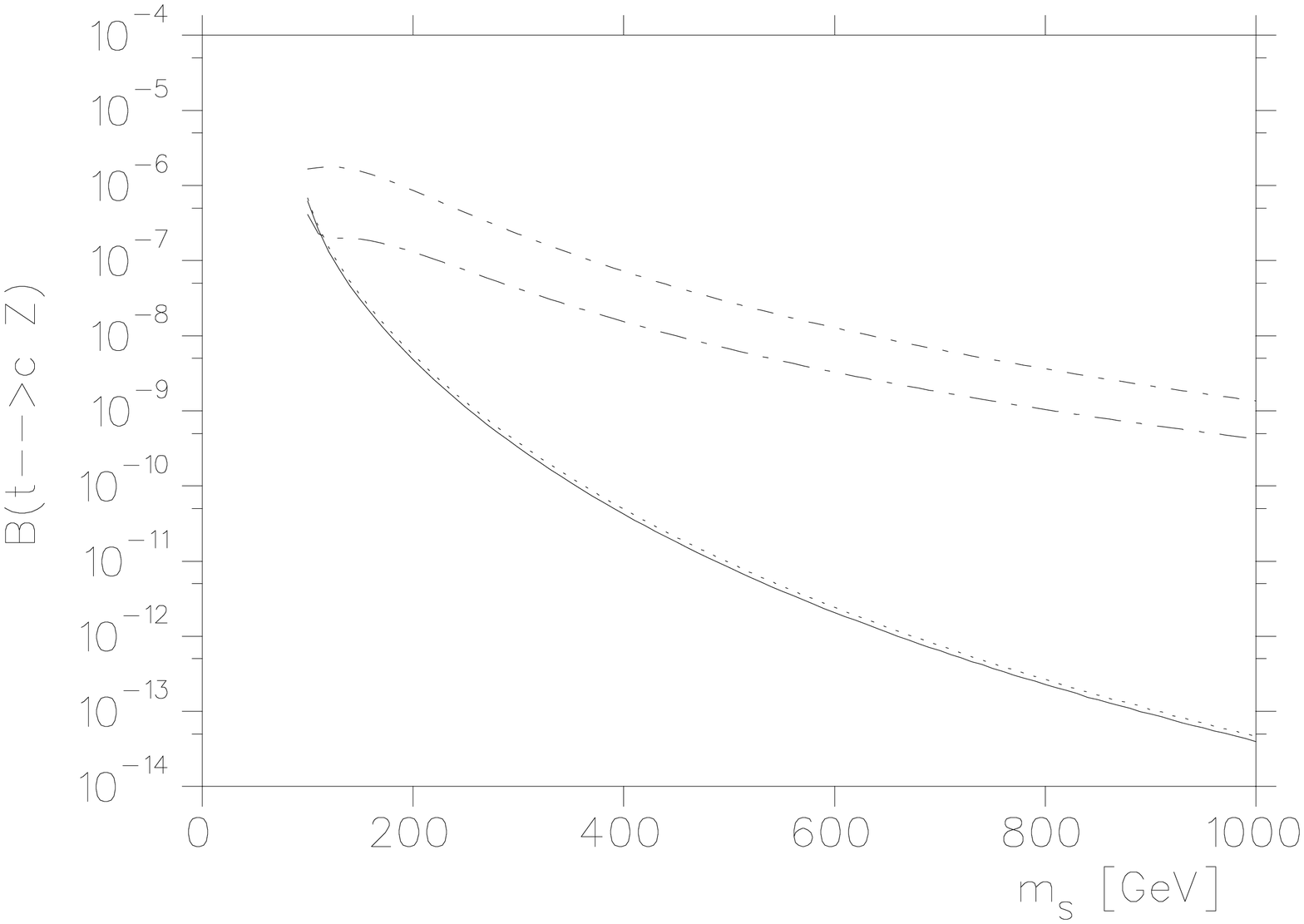}}
\end{center}
\caption{The ratio $\Gamma_S/\Gamma_W$\ of the
the top quark decay
into a charm quark and $Z^0$\ boson as a function of the
scalar mass $m_S$. The gluino mass was taken to be
100 GeV.
The solid line is
the unphysical case with no mixing ($\mu=0=A_{\rm top}$) and
$\tan\beta=10$, the dotted line the same case when flavour
changing $g-q-\tilde q$\ in the right-handed sector is included. 
The other cases are with mixing
($A_{\rm top}=m_S$).
The dashed--dotted ones with $\mu=500$\ GeV and $\tan\beta=10$.
The shorter ones are with flavour changing in both sectors.}
\label{Fig. 1}
\end{figure}

In Fig. 2 we consider the same cases as
in Fig. 1 but for $B(t\rightarrow c g)$. 
As before without mixing the results remain almost
the same whether or not flavour changing in the
right-handed sector is included. However when mixing
is taken into account the results are changed drastically
up to 7 oders of magnitude for large values
of the scalar mass $m_S$\ when flavour changing is considered
in both sectors, compared with the case where flavour changing
occurs only in the left-handed sector.

\begin{figure}[hbtp]
\begin{center}
\mbox{\epsfxsize=144mm\epsffile{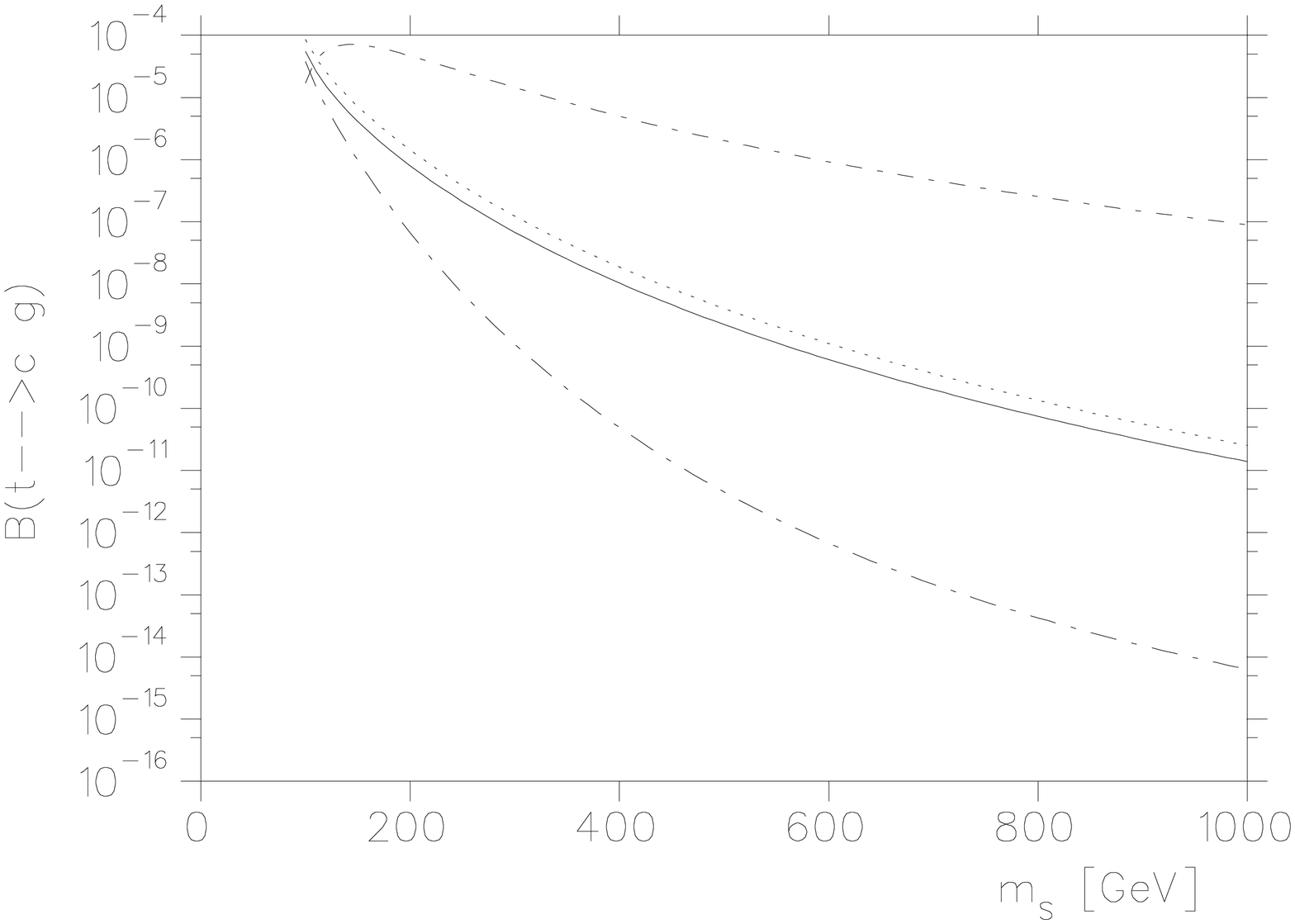}}
\end{center}
\caption{The same as Fig. 1  but for the
decay of the top quark into a charm quark and a gluon.}
\label{Fig. 2}
\end{figure}

In Fig. 3  we consider the branching ratio $B(t\rightarrow c \gamma)$
As in the cases before, without mixing there is almost no difference
between the results with flavour changing only in the left-handed
sector or in both sectors. As in Fig. 2 the results are changed
drastically, up to 6-7 orders of magnitude for large values
of the scalar mass, 
when mixing is taken into account and flavour changing
is considered in the left- and right-handed sector,
compared with the case where flavour changing occurs only in
left-handed sector.

A further important consequence is that the GIM-like supression where the
contribution of the top quark exactly cancels the contribution from the
c-quark is pushed to much smaller values of the scalar mass $m_S$. 
We have tried many different combinations of $\mu$ and 
$m_{\tilde g}$ and the cancellation is always pushed to smaller
values of the scalar mass. 

\begin{figure}[hbtp]
\begin{center}
\mbox{\epsfxsize=144mm\epsffile{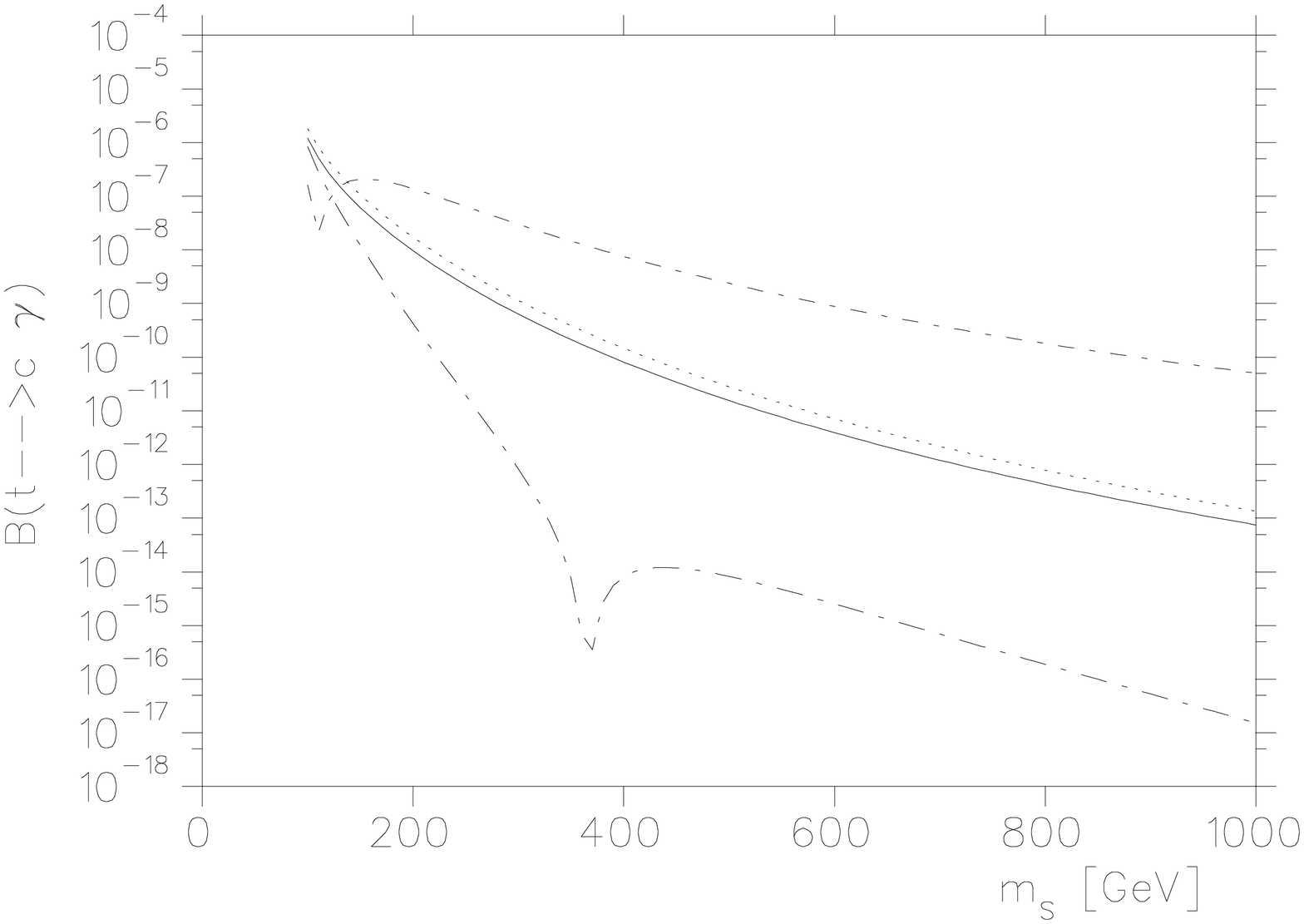}}
\end{center}
\caption{The same as Fig. 1  but for the
decay of the top quark into a charm quark and a photon.}
\label{Fig. 3}
\end{figure}

\section{\bf Conclusions}
\label{concl}

In this paper we presented the supersymmetric QCD 1-loop correction
to the flavour changing decay rate $t\rightarrow cV$.
We included flavour changing $g-q-\tilde q$\ couplings in
the left- and right-handed sector, thus extending the 
previous analysis of ~\cite{chk}, where flavour changing
was only considered in the left-handed sector.
We have shown that the results remain almost the same when
mixing of the scalar top quark is neglected. This remains
true for the $t\rightarrow c Z$\ decay rate even when mixing is
included. However the results are changed drastically, up to
7 orders of magnitude for the decay rates $t\rightarrow c g$\
and $t\rightarrow c\gamma$\ when mixing of the scalar top
quark is included and flavour changing couplings are taken
in both sectors. Furthermore in the $t\rightarrow c\gamma$\ 
decay mode the GIM-like cancellation of the scalar top and charm
quarks is pushed to much smaller values of the scalar mass $m_S$.

\hfill\break\vskip.1cm\noindent

{\bf Note}: While completing this work we have seen a paper
by an Italian group ~\cite{ita}, where the same processes
were considered. Their statement is that the SUSY mixing
angle between the second and the third generation
($K_{23}=\varepsilon$) has been over-estimated by at least
one order of magnitude in our first paper ~\cite{chk}.
There and in this present paper 
we took $\varepsilon$\ as a free
parameter and have taken it pretty large following the spirit of
former papers. From eq.(\ref{decrate}) it is obvious that the results are
diminished drastically if smaller values are taken for $\varepsilon$.
However the authors of ~\cite{ita} showed that relaxing the
universitality constraints on soft SUSY mass breaking terms of the
off-diagonal squark masses between $\tilde c$\ and $\tilde t$\
reintroduces a large $\varepsilon$, that is a large mixing angle
between $\tilde c$\ and $\tilde t$.

They also find a difference
in the result for the amplitude which can be traced
back to the ommission in ~\cite{chk} of the diagrams
involving a helicity flip in the gluino line, which
dominate the branching ratios when the gluino mass gets large.
However in ~\cite{chk} we considered flavour changing only
in the left-handed sector as is usually done in the MSSM 
and therefore no gluino helicity flip was possible,
that is no term proportional to the gluino mass is introduced.
In the present work, we also took into account flavour changing
in the right-handed sector and as a consequence the 
mentioned effect occurs, which is expressed by the new
terms proportional to the gluino mass in eq.(\ref{geneq}). 

\section{\bf Acknowledgements}

This work was funded by NSERC of Canada and les Fonds FCAR du Qu\'ebec.

\hfill\break

\end{document}